\documentclass[english,prl, reprint, superscriptaddress, showpacs]{revtex4-1}
\usepackage[T1]{fontenc}
\usepackage{amsmath}
\usepackage{amssymb}
\usepackage{graphicx}

\usepackage{babel}
\begin{document}

\title{An analytical limitation for time-delayed feedback control in autonomous
systems}

\author{Edward W. Hooton}

\affiliation{School of Mathematical Sciences, University College Cork, Ireland}

\author{Andreas Amann}

\email{a.amann@ucc.ie}

\affiliation{School of Mathematical Sciences, University College Cork, Ireland}

\affiliation{Tyndall National Institute, University College Cork, Lee Maltings, Cork, Ireland}
\begin{abstract}
We prove an analytical limitation on the use of time-delayed feedback
control for the stabilization of periodic orbits in autonomous systems.
This limitation depends on the number of real Floquet multipliers
larger than unity, and is therefore similar to the well-known odd
number limitation of time-delayed feedback control. Recently, a two-dimensional
example has been found, which explicitly demonstrates that the unmodified
odd number limitation does not apply in the case of autonomous systems.
We show that our limitation correctly predicts the stability boundaries
in this case. 
\end{abstract}


\maketitle
When chaotic systems started to get wider scientific attention during
the 1960s, chaos was considered to be a mathematically interesting
concept with little practical applications. This changed dramatically
in the 1990s when Ott, Grebogi and Yorke \citep{OTT90} introduced
a method to stabilize Unstable Periodic Orbits (UPOs) within the chaotic
attractor using small perturbations. Since then, the subject of chaos
control has been vigorously developed \citep{SCH07d,BOC00}. 

One simple method to stabilize a particular UPO within a chaotic attractor
is via the time-delayed feedback control due to Pyragas \citep{PYR92}.
Because no detailed knowledge of the chaotic system or its attractor
is required, this method proved to be easy to implement and widely
applicable \citep{PYR93,JUS97,SUK97,ROS04a,SCH06g,SIE08,TLI09,ENG10}.
However, it was claimed by Nakajima \citep{NAK97} that the time-delayed
feedback control is not able to stabilize a UPO with an odd number
of real Floquet multipliers  larger than unity. While this odd number
limitation was proved in \citep{NAK97} for the case of hyperbolic
UPOs in non-autonomous systems, it was also stated that the same restriction
should apply for the autonomous case {}``with a slight revision''
(footnote 2 of \citep{NAK97}). Over the following years the odd number
limitation was used by many researchers, and it seemed to be supported
by experimental and numerical evidence even for autonomous systems,
although in this case a strict proof was missing. Recently, Fiedler
et al. \citep{FIE07} discovered a UPO in an autonomous two-dimensional
system, which has precisely one Floquet multiplier larger than one,
and can be stabilized by the time-delay feedback control scheme. This
directly refuted the common belief that the odd number limitation
is also valid for systems without explicit time-dependence. Autonomous
systems are by far the most dominating type of systems considered
in nonlinear science, and time-delayed feedback control is one of
the most practical methods for stabilizing (or destabilizing) periodic
orbits. Therefore any limitation on the use of time-delayed feedback
control is not only important from an academic point of view, but
also has practical implications for the many applications of time-delayed
feedback in real-world systems. 

In this Letter we give an analytical condition under which the
time-delayed feedback control is not successful in autonomous
systems. Similarly to the odd number limitation, this condition
involves the number of real Floquet multipliers larger than unity, but
it is now modified by a term which takes the action of the control
force in the direction of the periodic orbit into account.  We will
also connect this modification to the response of the system to
changes in the delay time.  Our proof follows to a large extent the
proof of the original odd number limitation given in \citep{NAK97} but
now implements the necessary modification for the autonomous case.  As
a first application we show that our limitation correctly reproduces
the boundaries of stability for the two dimensional system studied in
\citep{FIE07,JUS07,FLU11}, which originally served as a counter
example of the unmodified odd number limitation.

Let us start with an uncontrolled dynamical system $\dot{x}(t)=f\left(x(t)\right)$
with $x\left(t\right)\in\mathbb{R}^{n}$ and $f:\mathbb{R}^{n}\to\mathbb{R}^{n}$
and implement the time-delayed feedback control in the form 
\begin{equation}
\dot{x}(t)=f\left(x(t)\right)+K\left[x(t-\tau)-x(t)\right],\label{eq:dynsys}
\end{equation}
where $K$ is an $n\times n$ control matrix, and $\tau$ is a positive
parameter. If the uncontrolled system has a $\tau$-periodic solution
$x^{*}\left(t\right)=x^{*}\left(t+\tau\right),$ then the form of
\eqref{eq:dynsys} implies that $x^{*}\left(t\right)$ is also a solution
of \eqref{eq:dynsys} for any choice of the control matrix $K$.

In order to assess the stability of the periodic orbit $x^{*}\left(t\right)$
in the controlled case, it is convenient to first introduce the the
fundamental matrix $\Phi\left(t\right)$ for the \emph{uncontrolled}
system as the solution of the initial value problem 
\begin{equation}
\dot{\Phi}\left(t\right)=Df\left(x^{*}\left(t\right)\right)\Phi\left(t\right); 
\qquad\Phi\left(0\right)=\mathbb{I},\label{eq:Phidef}
\end{equation}
where $Df\left(x^{*}\left(t\right)\right)$ denotes the Jacobian of
$f$ evaluated at $x^{*}\left(t\right)$, and $\mathbb{I}$ is the
$n\times n$ identity matrix. The generalized eigenvalues $\left\{ \mu_{1},\ldots,\mu_{n}\right\} $
of $\Phi\left(\tau\right)$ are the Floquet multipliers associated
with the periodic orbit $x^{*}\left(t\right).$ We also define the
matrix $W\left(t\right)=\left(v_{1}\left(t\right),\ldots,v_{n}\left(t\right)\right)$
such that its $k$th column $v_{k}\left(t\right)\in\mathbb{C}^{n}$
is given by $v_{k}\left(t\right)=\Phi\left(t\right)v_{k}\left(0\right)$
and the set $\left\{ v_{1}\left(0\right),\ldots,v_{n}\left(0\right)\right\} $
is a Jordan basis of generalized eigenvectors of $\Phi\left(\tau\right).$
For each $t$, the set $\left\{ v_{1}\left(t\right),\ldots,v_{n}\left(t\right)\right\} $
provides a local (but in general not $\tau$-periodic) basis at the
position $x^{*}\left(t\right)$ along the orbit. Since we consider
an autonomous system we also observe that 
$\dot{x}^{*}\left(0\right)=\Phi\left(\tau\right)\dot{x}^{*}\left(0\right)$,
i.e. one of the Floquet multipliers is equal to unity. It is therefore
convenient to choose $v_{1}\left(t\right)=\dot{x}^{*}\left(t\right)$
and $\mu_{1}=1$. By defining 
\begin{equation}
\hat{K}\left(t\right)=\left[W\left(t\right)\right]^{-1}KW\left(t\right)\label{eq:Khat}
\end{equation}
we transform the control matrix to this local basis. As we will see
in the following the $\left(1,1\right)$ component of the matrix $\hat{K}\left(t\right)$,
which we denote by $\hat{K}_{11}\left(t\right)$, plays a decisive
role in assessing the stability of the controlled orbit $x^{*}\left(t\right).$
Some intuition for the quantity $\hat{K}_{11}\left(t\right)$ can
be obtained if we expand the result of applying the control matrix
to $\dot{x}^{*}\left(t\right)$ in the local basis via 
\begin{equation}
K\dot{x}^{*}\left(t\right)= \hat{K}_{11}\left(t\right)\dot{x}^{*}\left(t\right) 
+\sum_{k=2}^{n}\hat{K}_{k1}\left(t\right)v_{k}\left(t\right).\label{eq:k11}
\end{equation}
In loose terms we can therefore interpret the quantity $\hat{K}_{11}\left(t\right)$
as the action of the control matrix $K$ projected in the tangential
direction of the orbit at time $t.$ Note that $\hat{K}_{11}\left(t\right)$
is well defined and in particular not affected by any reordering or
rescaling of the $v_{k}\left(t\right)$ for $k\geq2$. Using this
definition of $\hat{K}_{11}\left(t\right)$ we are now in a position
to formulate the main result. 

\textbf{Theorem:} Let $x^{*}\left(t\right)$ be a $\tau$-periodic
orbit of \eqref{eq:dynsys} which for $K=0$ possesses $m$ real Floquet
multipliers larger than unity and precisely one Floquet multiplier
equal to unity. Then $x^{*}\left(t\right)$ is an unstable solution
of the time-delayed system \eqref{eq:dynsys} if the condition 
\begin{equation}
\left(-1\right)^{m}\left(1+\int_{0}^{\tau}\hat{K}_{11}\left(t\right)dt\right)<0,\label{eq:condition}
\end{equation}
is fulfilled. Here $\hat{K}_{11}\left(t\right)$ is defined as in
\eqref{eq:k11}.

Before we proceed with the proof of the theorem, we briefly discuss
its significance and reformulate it in a way which is more useful for 
practical applications. The theorem provides an analytical limitation on
the use of time-delayed feedback control, and states that time-delayed
feedback can \emph{only} successfully stabilize a periodic orbit,
if the condition \eqref{eq:condition} is violated. We stress however
that the converse is not implied by the theorem, i.e. a violation
of \eqref{eq:condition} alone does \emph{not} guarantee that time-delayed
feedback will successfully stabilize a given periodic orbit. The theorem
is only applicable to periodic orbits with exactly one Floquet multiplier
equal to one. 

In practice the integral over the matrix element
$\hat{K}_{11}\left(t\right)$ in \eqref{eq:condition} is difficult to
perform, even if the system is analytically known.  A practically
useful reformulation of condition \eqref{eq:condition} can be obtained
from studying the response of the system to changes in the delay
time. We consider a variant of the system \eqref{eq:dynsys}
\begin{equation}
\dot{x}(t)= 
f\left(x(t)\right)+K\left[x(t-\hat{\tau})-x(t)\right],\label{eq:dynsystauhat}
\end{equation}
where the delay time $\hat{\tau}$ is slightly different from the
period $\tau$ of the uncontrolled orbit.  For $\hat{\tau}$
sufficiently close to $\tau$ the system \eqref{eq:dynsystauhat} will
possess a (possibly unstable) induced periodic orbit
$\tilde{x}^{*}\left(t\right)$ with period
$\tilde{\tau}\left(\hat{\tau}\right)$.  In general $\tilde{\tau}$ is
different from both $\tau$ and $\hat{\tau}$, however one can show
that the period of the induced orbit is connected with the matrix element 
$\hat{K}_{11}\left(t\right)$ via 
\begin{equation}
\lim_{\hat{\tau}\to \tau} \frac{\tilde{\tau}\left(\hat{\tau}\right)-\tau}{\hat{\tau} -\tilde{\tau}\left(\hat{\tau}\right)}= \int_{0}^{\tau}\hat{K}_{11}\left(t\right)dt.
\end{equation}
From the condition \eqref{eq:condition} of our theorem it then follows
that $x^{*}\left(t\right)$ is an unstable solution of the system
\eqref{eq:dynsys} if the condition
\begin{equation}
(-1)^m \lim_{\hat{\tau}\to\tau}  \frac{\hat{\tau}-\tau}{\hat{\tau}-\tilde{\tau}\left(\hat{\tau}\right)}<0 \label{eq:condtautilde}
\end{equation}
holds.  Since condition \eqref{eq:condtautilde} only requires the
knowledge of the period of the induced orbit $\tilde{\tau}$ as a
function of the delay time $\hat{\tau}$, it is often more convenient in practice
than the equivalent but more technical condition \eqref{eq:condition}.

The proof of the theorem uses many
ideas from \citep{NAK97} but now takes particular consideration of
the autonomous case. The essential tools are the two
functions $F\left(\nu\right)$ and $G\left(\nu\right)$ defined by
\citep{NAK97}
\begin{eqnarray}
G\left(\nu\right) & = & \det\left[\nu\mathbb{I}-\Phi\left(\tau\right)\right]\label{eq:Gdef}\\
F\left(\nu\right) & = & \det\left[\nu\mathbb{I}-\Psi_{\nu}\left(\tau\right)\right]\label{eq:Fdef}
\end{eqnarray}
where $\Psi_{\nu}\left(t\right)$ solves the initial value value problem
\begin{eqnarray}
\dot{\Psi}_{\nu}\left(t\right) & = & \left[Df\left(x^{*}\left(t\right)\right)+\left(\nu^{-1}-1\right)K\right]\Psi_{\nu}\left(t\right),\label{eq:Psidef}\\
\Psi_{\nu}\left(0\right) & = & \mathbb{I}.\nonumber 
\end{eqnarray}
By direct differentiation it can be verified that the solution of
\eqref{eq:Psidef} can also be expressed as \citep{NAK97} 
\begin{equation}
\Psi_{\nu}\left(t\right)=\Phi\left(t\right)\left[\mathbb{I}+\left(\nu^{-1}-1\right)\int_{0}^{t}\Phi^{-1}\left(u\right)K\Psi_{\nu}\left(u\right)du\right].\label{eq:reprPsi}
\end{equation}

We first show the following lemma:

\textbf{Lemma}: If for a given $\tau$-periodic orbit $x^{*}\left(t\right)$
with precisely one Floquet multiplier equal to unity the condition
$F'\left(1\right)<0$ holds, then $x^{*}\left(t\right)$ is an unstable
solution of the time-delayed system \eqref{eq:dynsys}. 

\textbf{Proof of the Lemma}: From \eqref{eq:Psidef} it follows that
$\Psi_{\nu}\left(t\right)$ is bounded in the limit of $\nu\to+\infty,$
and therefore \eqref{eq:Fdef} implies that $\lim_{\nu\to+\infty} F\left(\nu\right)=+\infty$.
In an autonomous system we know
that $\mu_{1}=1$ is an eigenvalue of $\Phi\left(\tau\right)$ and
it therefore follows from \eqref{eq:Gdef} that $G\left(1\right)=0$.
But since from \eqref{eq:Psidef} it follows that $\Psi_{1}\left(t\right)=\Phi\left(t\right)$
we also have $F\left(1\right)=G\left(1\right)=0$. Thus $F\left(\nu\right)$
is a continuous function, which vanishes at $\nu=1$, has a negative
slope at $\nu=1,$ and diverges to $+\infty$ for large $\nu$. By
the intermediate value theorem there exists at least one $\nu_{c}>1$
with $F\left(\nu_{c}\right)=0$. From \eqref{eq:Fdef} it then follows
that there exists a vector $w_{c}\left(0\right)\in\mathbb{R}^{n}$
with $\Psi_{\nu_{c}}\left(\tau\right)w_{c}\left(0\right)=\nu_{v}w_{c}\left(0\right)$
and we can define $w_{c}\left(t\right)=\Psi_{\nu_{c}}\left(t\right)w_{c}\left(0\right)$.
Then $w_{c}\left(t\right)$ is a growing solution of the linearized
equation $\dot{w}_c\left(t\right) =
Df\left(x^{*}\left(t\right)\right)w_c\left(t\right) 
+ K \left[w_{c}\left(t-\tau\right)-w_{c}\left(t\right)\right] $ and therefore the
original orbit $x^{*}\left(t\right)$ is an unstable solution of the
time-delayed system \eqref{eq:dynsys}. This completes the proof of the
lemma.

\textbf{Proof of the Theorem: }To complete the proof of the theorem,
it now remains to show that the condition \eqref{eq:condition} implies
that $F'\left(1\right)<0$. Under this condition the above lemma then
implies that the orbit $x^{*}\left(t\right)$ is unstable and thereby
proves the theorem. Since $F\left(1\right)=0$ we can write 
$F'\left(1\right)=\lim_{\epsilon\to0}F\left(1+\epsilon\right)/\epsilon.$
To assess the sign of $F'\left(1\right)$ it is therefore necessary
to evaluate $F\left(1+\epsilon\right)$ up to first order in $\epsilon$.
According to \eqref{eq:Fdef} we can achieve this by first evaluating
$\Psi_{\nu}\left(t\right)$ at $\nu=1+\epsilon$. Using the representation
\eqref{eq:reprPsi} we write 
\begin{eqnarray*}
\Psi_{1+\epsilon}\left(t\right) & = & \Phi\left(t\right)\left[\mathbb{I}-\frac{\epsilon}{1+\epsilon}\int_{0}^{t}\Phi^{-1}\left(u\right)K\Psi_{1+\epsilon}\left(u\right)du\right]\\
 & = & \Phi\left(t\right)\left[\mathbb{I}-\epsilon\int_{0}^{t}\Phi^{-1}\left(u\right)K\Phi\left(u\right)du\right]+O\left(\epsilon^{2}\right).
\end{eqnarray*}
Then to first order of $\epsilon$ we obtain from \eqref{eq:Fdef}
\begin{eqnarray}
F\left(1+\epsilon\right) & = & \det\left[M^{0}+\epsilon M^{1}\right]\label{eq:detform}
\end{eqnarray}
where we have defined the two matrices 
\begin{equation}
M^{0}  =  \mathbb{I}-\Phi\left(\tau\right);\,
M^{1}  =  \mathbb{I}+\Phi\left(\tau\right)\int_{0}^{t}\Phi^{-1}\left(u\right)K\Phi\left(u\right)du.\nonumber
\end{equation}
Using the previously defined matrix $W\left(0\right)$ we transform the argument of the determinant
in \eqref{eq:detform} as
\begin{eqnarray}
 &  & F\left(1+\epsilon\right)=\det\left[W\left(0\right)^{-1}\left(M^{0}+\epsilon M^{1}\right)W\left(0\right)\right]\label{eq:detF}\\
 & = & \det\left[\mathbb{I}-\hat{\Phi}\left(\tau\right)+\epsilon\left(\mathbb{I}+\hat{\Phi}\left(\tau\right)\int_{0}^{\tau}\hat{K}\left(u\right)du\right)\right]\label{eq:detF2}
\end{eqnarray}
where we have used $W\left(0\right)^{-1}\Phi^{-1}\left(u\right)K\Phi\left(u\right)W\left(0\right)=W\left(u\right)^{-1}KW\left(u\right)=\hat{K}\left(u\right).$
Using the Jordan normal form of $\hat{\Phi}\left(\tau\right)$ and
the fact that $\mu_{1}=1$ we find 

\begin{equation}
\mathbb{I}-\hat{\Phi}\left(\tau\right)=\left(\begin{array}{ccccc}
0 & 0 & 0 & \cdots & 0\\
0 & 1-\mu_{2} & * & \ddots & \vdots\\
\vdots & \ddots & \ddots & \ddots & 0\\
\vdots &  & \ddots & \ddots & *\\
0 & \cdots & \cdots & 0 & 1-\mu_{n}
\end{array}\right)\label{eq:matzero}
\end{equation}
where the entries indicated by $*$ can be either $0$ or $-1$. Since
all entries in the first column and first row of the matrix $\mathbb{I}-\hat{\Phi}\left(\tau\right)$
vanish, the only contribution to the determinant \eqref{eq:detF2}
up to first order in $\epsilon$ is given by 
\begin{eqnarray*}
F\left(1+\epsilon\right) & = & \epsilon\left(\mathbb{I}+\hat{\Phi}\left(\tau\right)\int_{0}^{\tau}\hat{K}\left(u\right)du\right)_{\left(1,1\right)}\prod_{k=2}^{n}\left(1-\mu_{k}\right)\\
 & = & \epsilon\left(1+\int_{0}^{\tau}\hat{K}_{11}\left(u\right)du\right)\prod_{k=2}^{n}\left(1-\mu_{k}\right).
\end{eqnarray*}
The product $\prod_{k=2}^{n}\left(1-\mu_{k}\right)$ does not vanish,
since it was assumed that $\mu_{1}=1$ is the only Floquet multiplier
equal to unity. Each real Floquet multiplier larger than unity contributes
a negative factor to this product, while pairs of complex conjugated
Floquet multipliers or real Floquet multipliers smaller than unity
do not change its sign. We can therefore write $\mbox{sgn}\left(\prod_{k=2}^{n}\left(1-\mu_{k}\right)\right)=\left(-1\right)^{m}$
and conclude that 
\[
F'\left(1\right)<0\,\Leftrightarrow\,\left(-1\right)^{m}\left(1+\int_{0}^{\tau}\hat{K}_{11}\left(u\right)du\right)<0.
\]
This means that condition \eqref{eq:condition} implies a negative
$F'\left(1\right)$ and it then follows from the lemma that $x^{*}\left(t\right)$
is unstable. This concludes the proof of the theorem.

Let us now compare our theorem and its proof with the proof of the
original odd number limitation (theorem 2 in \citep{NAK97}), which
states that a \emph{hyperbolic} UPO of a non-autonomous system with an
odd number $m$ of real Floquet multipliers larger than unity can not
be stabilized using time-delayed feedback control. We stress that the
term \emph{hyperbolic orbit} in the context of a non-autonomous system
means that the orbit has \emph{no} Floquet multipliers equal to
unity. In contrast, for an autonomous system the term \emph{hyperbolic
  orbit} denotes an orbit with \emph{precisely one} Floquet multiplier
equal to unity \citep{KUZ04}. Therefore any hyperbolic orbit in the
autonomous system becomes a non-hyperbolic orbit in the associated
non-autonomous system. The proof in \citep{NAK97} however makes
explicit use of the fact that \emph{all} Floquet multipliers differ
from one, and is therefore only correct if the term \emph{hyperbolic}
is understood in the context of non-autonomous systems.  

Let us now study condition \eqref{eq:condtautilde} for odd $m$.  If we
increase the delay time to $\hat{\tau}>\tau$ the system will respond
with a period $\tilde{\tau}$ of the induced orbit.  One might now be
tempted to assume that the period of the induced orbit should always
be less than the delay time, i.e.  $\tilde{\tau}< \hat{\tau}$. This is
however not the case, and it is possible to find dynamical systems,
which respond to an increased delay time with an induced period which
is even bigger than the delay time itself. In this case
\eqref{eq:condtautilde} is violated and it might be possible to
stabilize the corresponding periodic orbit using time-delayed feedback
control.  This consideration leads to an important practical
consequence for the design of a successful control scheme for a UPO
with an odd number of Floquet multipliers larger than 1. The control
term needs to be constructed in such a way that for increasing delay
time the period of the induced orbit grows faster than the delay time itself.

\begin{figure}
  \includegraphics[width=1\columnwidth]{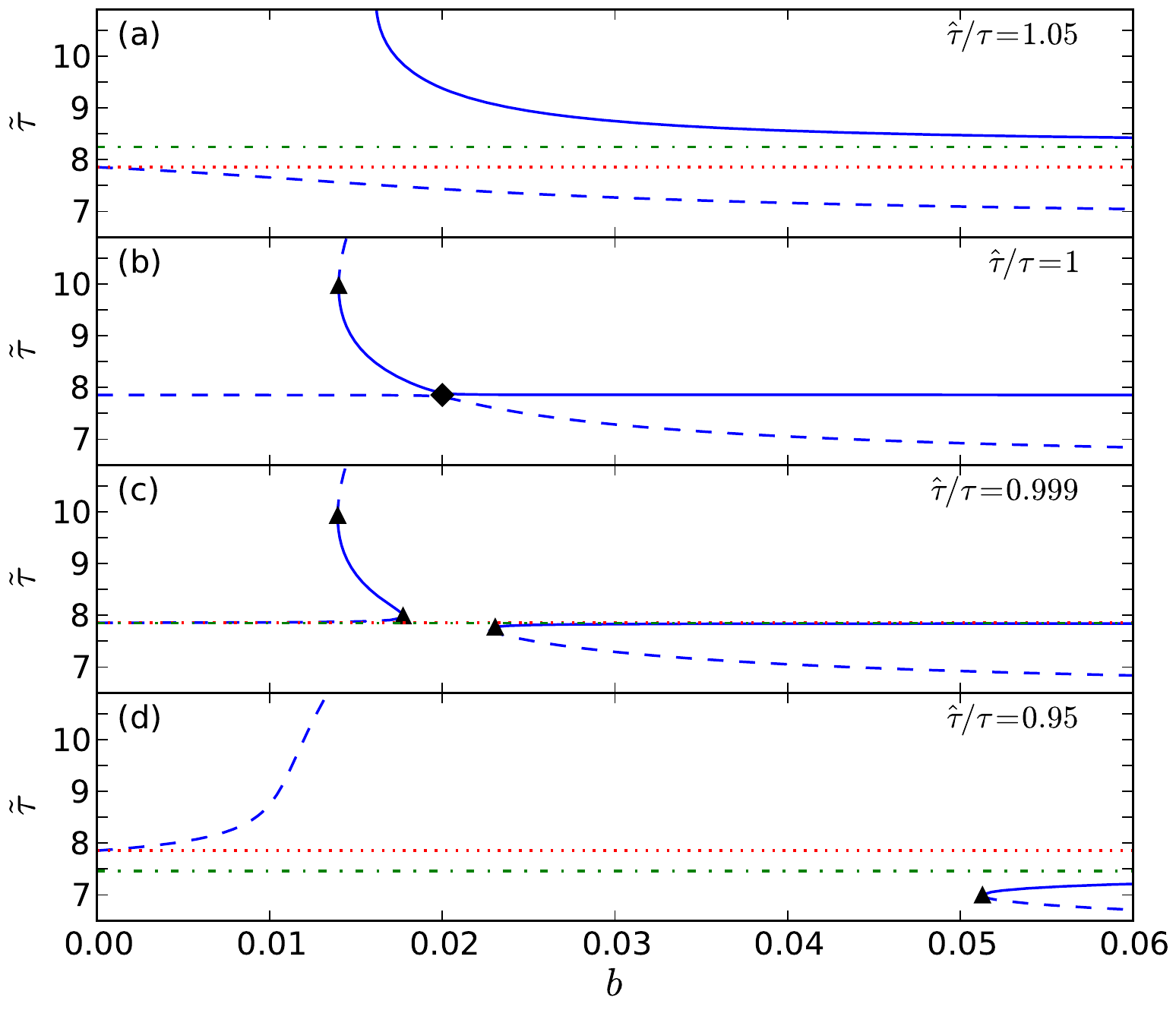}
  \caption{The periods $\tilde{\tau}$ of stable (solid lines)
    and unstable (dashed lines) periodic orbits of the system
    \eqref{eq:examplesys} as a function of $b$ for different values
    of $\hat{\tau}$.  Calculations were performed using
    Knut \cite{SZA09}. Triangles and diamond
    indicate saddle-node and transcritical bifurcations,
    respectively. The values of   $\tau$ and $\hat{\tau}$ are 
    indicated by the horizontal dotted (red) and dashed-dotted (green) 
    lines, respectively.  \label{fig:indorb}}
\end{figure}

The first autonomous example, where a UPO with odd $m$ was stabilized
using time-delayed feedback control was given in \citep{FIE07} and
provided a counter example which showed that the original odd number
limitation can not be applied to the autonomous case without modification.
It is therefore important to check that our conditions \eqref{eq:condition} and \eqref{eq:condtautilde} 
correctly handle this case. Let us consider the dynamical system
for $z\left(t\right)\in\mathbb{R}^{2}$ given by \citep{FIE07,JUS07}
\begin{eqnarray}
\dot{z}\left(t\right) & = & \left(\begin{array}{cc}
\left(\left|z\left(t\right)\right|^{2}-R^{2}\right) & -\left(1+\gamma\left|z\left(t\right)\right|^{2}\right)\\
\left(1+\gamma\left|z\left(t\right)\right|^{2}\right) & \left(\left|z\left(t\right)\right|^{2}-R^{2}\right)
\end{array}\right)z\left(t\right)\nonumber \\
 &  & +b\left(\begin{array}{cc}
\cos\beta & -\sin\beta\\
\sin\beta & \cos\beta
\end{array}\right)\left[z\left(t-\hat{\tau}\right)-z\left(t\right)\right],\label{eq:examplesys}
\end{eqnarray}
with the main bifurcation parameters $\hat{\tau}>0$ and $b\geq 0$. For
the remaining parameters we choose $R^2=0.02$, $\gamma=-10$ and
$\beta=\pi/4$.  For $b=0$ we find a periodic orbit
$z^{*}\left(t\right)=R\left(\cos\left(2\pi
    t/\tau\right),\sin\left(2\pi t/\tau\right)\right)^{T}$ with period
$\tau=2\pi/\left(\gamma R^2 + 1\right)>0$.  A short calculation shows
that the two Floquet multipliers are given by $\mu_{1}=1$ and
$\mu_{2}=\exp\left(2R^{2}\tau\right)>1$ which implies $m=1$.  

For $\hat{\tau}=\tau$ and increasing $b$ the unstable orbit
$z^{*}\left(t\right)$ is stabilized via a transcritical bifurcation at
a critical value $b_c$ \cite{FIE07}.  This scenario is shown in
Fig.~\ref{fig:indorb}(b), where the transcritical bifurcation is
indicated by a diamond.  We now change the delay time $\hat{\tau}$ and
study the period $\tilde{\tau}$ of the induced orbit which
continuously connects to the orbit $z^{*}\left(t\right)$. For
$\hat{\tau}>\tau$, we observe that the transcritical bifurcation
evolves into an avoided crossing of two branches, as shown in
Fig.~\ref{fig:indorb}(a).  For $b<b_c$ the periodic orbit
$z^{*}\left(t\right)$ at $\hat{\tau}=\tau$ evolves into an orbit with
period $\tilde{\tau}<\hat{\tau}$. Therefore condition
\eqref{eq:condtautilde} is fulfilled, and our theorem guarantees that
the orbit $z^{*}\left(t\right)$ is unstable for $\hat{\tau}=\tau$ and
$b<b_c$.  For $\hat{\tau}>\tau$ and $b>b_c$ we observe that
$\tilde{\tau} > \hat{\tau}$.  This means that as we increase the delay
time, the period of the induced orbit becomes even larger than the new
delay time.  In this intuitively unusual case the condition
\eqref{eq:condtautilde} is violated.  Thus our theorem does not apply
for $b>b_c$ and stabilization is possible.  Similar considerations
apply for the case $\hat{\tau}<\tau$ (Fig.~\ref{fig:indorb}(c)+(d))
where the transcritical bifurcation evolves into a pair of saddle-node
bifurcations.  Again condition \eqref{eq:condtautilde} is only fulfilled for  $b<b_c$.

For this simple example we can also explicitly calculate
$\hat{K}_{11}\left(t\right)=b\left(\cos\beta+\gamma\sin\beta\right)$.
For the possibility of successful stabilization we need to violate the condition
\eqref{eq:condition}, which leads to the necessary condition
\begin{equation}
b\left(\cos\beta+\gamma\sin\beta\right)\tau<-1.\label{eq:cond_example}
\end{equation}
This agrees with the previous stability condition given in
\citep{JUS07} and the location of the transcritical bifurcation in
Fig.~\ref{fig:indorb}(b).  

\begin{figure}
  \includegraphics[width=1\columnwidth]{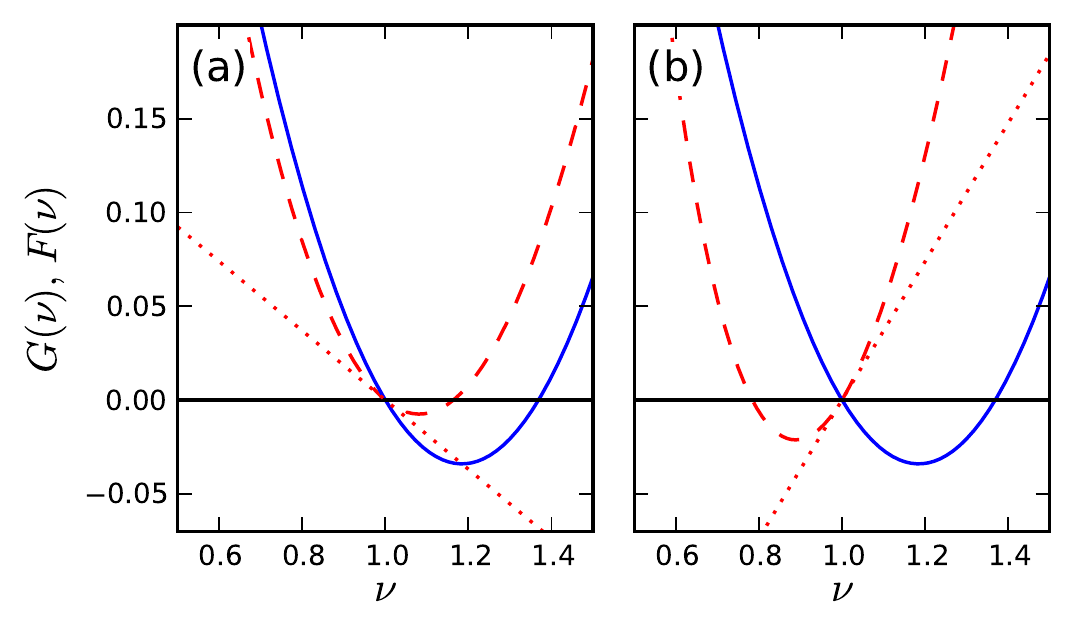}\caption{The functions
    $G\left(\nu\right)$ (solid blue lines) and $F\left(\nu\right)$ (dashed red
    lines) and the slopes $F'\left(1\right)$ (dotted red lines) for system
    \eqref{eq:examplesys} for $b=0.01$ (a) and $b=0.04$
    (a) and with $\hat{\tau}=\tau$. \label{fig:GF}}
\end{figure}

It is also illustrative to study the functions $F\left(\nu\right)$
\eqref{eq:Fdef} and $G\left(\nu\right)$ \eqref{eq:Fdef} for the
current example.  In Fig.~\ref{fig:GF}(a) the corresponding plots are
shown for $b<b_c$. In this case we observe that $F\left(1\right)=0$
and the slope $F'\left(1\right)$ is negative. Therefore the function
$F\left(\nu\right)$ needs to cross the zero axis at a point larger
than unity and the periodic orbit is unstable. In the case of $b>b_c$,
as shown in Fig.~\eqref{fig:GF}(b), the slope $F'\left(1\right)$ is
positive and the function $F\left(\nu\right)$ does not cross the zero
axis at values larger then unity. This is a necessary condition for
the stability of the periodic orbit under time-delayed feedback
control.

In conclusion, we have proved an analytical limitation on the use of
time-delayed feedback control in autonomous systems. This limitation
depends on the number of real Floquet multipliers larger than unity,
and on the properties of the induced orbits as the delay time is
varied. While the limitation is valid for arbitrary dimensions, we
have demonstrated its usefulness in a well studied two-dimensional
system, for which the original odd number limitation does not
apply. The knowledge of this limitation will provide an important
guidance for the design of time-delayed feedback implementations in
practical applications.

This work was supported by Science Foundation Ireland under Grant
Number 09/SIRG/I1615. 

%

\end{document}